\documentclass[trackchanges]{aastex701}

\begin{document}

\title{Numerically stable equations for the orbital evolution of compact object binaries}

\author[orcid=0000-0002-6842-3021,sname='Briel']{Max M. Briel}
\affiliation{Département d’Astronomie, Université de Genève, Chemin Pegasi 51, CH-1290 Versoix, Switzerland}
\affiliation{Gravitational Wave Science Center (GWSC), Université de Genève, CH-1211 Geneva, Switzerland}
\email[show]{max.briel@gmail.com}

\author[orcid=0000-0001-5261-3923,sname='Andrews']{Jeff J. Andrews}
\affiliation{Department of Physics, University of Florida, 2001 Museum Rd, Gainesville, FL 32611, USA}
\affiliation{Institute for Fundamental Theory, 2001 Museum Rd, Gainesville, FL 32611, USA}
\email{jeffrey.andrews@ufl.edu}

\begin{abstract}
The orbital and eccentricity evolution for compact object binaries through gravitational wave emission first derived by Peters and Mathews are used extensively throughout the gravitational wave community for calculating the orbital evolution and merger time of compact binaries. While improved calculations of the binary merger time have been the focus of several investigations since, the orbital evolution has not received the same attention. As the equations lack a closed form solution, a numerical integrator is required, but standard methods typically break when the point of merger is overstepped. We present a rewrite of Peters' equations in $\ln$-space, which allows common numerical solvers to converge. This leads to a more numerically robust and computationally efficient method for evolving compact binaries due to gravitational wave emission, reducing the number of function evaluations by 60\% to 70\% in our tests.
\end{abstract}


\section{Introduction}

    The orbital evolution due to gravitational wave emission is crucial in understanding the population properties of compact object binaries. For this purpose, the orbital averaged equations derived by \citet{Peters+63} and \citet{Peters+64} for the evolution of orbital separation $a$ and eccentricity $e$ are used:
    
    \begin{equation} \label{eq:de_dt}
        \left< \frac{de}{dt} \right> =  -\frac{304}{15}\frac{G^3m_1m_2(m_1+m_2)}{c^5a^4(1-e^2)^{5/2}} e\left(1+\frac{121}{304}e^2 \right)
    \end{equation}
    
    \begin{equation} \label{eq:da_dt}
        \left<\frac{da}{dt}\right> = -\frac{64}{5} \frac{G^3 m_1 m_2 (m_1 + m_2)}{c^5 a^3 \left(1 - e^2\right)^{7/2}} \left(1 + \frac{73}{24} e^2 + \frac{37}{96} e^4\right),
    \end{equation}
    where $G$ is the gravitational constant, $c$ is the speed of light, and $m_1$ and $m_2$ are the masses of the binary components. We note that, although orbital evolution has been calculated to higher post-Newtonian orders, most notably by \citet{Junker+92}, in this work we only consider the lowest order form provided by Equations~\ref{eq:de_dt} and \ref{eq:da_dt}.
    
    When one is interested in the number of compact object mergers, such as for BBH mergers in the LIGO/Virgo/KAGRA detector network, only the merger time needs to be calculated. This time can be calculated by numerically integrating the expression provided by \citet{Peters+64}, applying hypergeometric functions \citep{Pierro+96, Pierro+02}, or by using the analytical approximations from e.g., \citet{Mandel+21} when computational time is essential.
    
    However, for non-merging gravitational wave sources, such as Galactic double white dwarf or binary neutron star sources, the differential equations need to be solved to find the orbital configuration of the system over time. 
    Evolving the system requires solving the coupled Peters equations up until a requested age or the system's merger. As the separation decreases ($a\rightarrow0$), the $1/a^3$ term in Equation~\ref{eq:da_dt} diverges, requiring a very small timestep to maintain numerical stability for $da/dt$. Furthermore, a singularity occurs at $a=0$: for small positive $a$, evaluating Equation~\ref{eq:da_dt} leads to large negative derivatives, while non-physical negative values for $a$ lead to large positive derivatives. Numerical integrators with adaptive step sizes struggle to converge near this singularity point. A similar behavior arises in Equation \ref{eq:de_dt} which contains a $1/a^4$ dependence.
    
    While this singularity can be avoided by stopping the integration before approaching too close to the moment of merger, one can, depending on the problem, miss important evolution that occurs at small separations. Furthermore, the different scales of the problems for different compact object types make a universal solution challenging. For instance, a white dwarf has a radius of ${\sim}7000$ km, while a neutron star has a radius of $\simeq$10 km. An ideal solution allows for the integration of systems with separations as wide as $\sim$AU and as small as $\sim$km\footnote{At separations of a few gravitational radii and for orbits with eccentricities approaching unity, higher-order post-Newtonian corrections to Equations~\ref{eq:de_dt} and \ref{eq:da_dt} are needed.}, requiring reasonable integration accuracy over a range spanning more than eight orders of magnitude.
    
\section{Transforming the Peters Equations}

    A first step towards a solution is rewriting Equations~\ref{eq:de_dt} and \ref{eq:da_dt} in a dimensionless form. This eliminates units as source of numerical error, and allows the solver to handle both stellar mass and supermassive compact object binaries. Following \citet{Andrews+19} we replace $\alpha = a/a_0$, where $a_0$ is the system's initial orbital separation and $\tau = t/t_0$ where: 
    \begin{equation}
        t_0 = \frac{5}{64}\frac{c^5a_0^4}{G^3(m_1+m_2)m_1m_2}.
    \end{equation}
    Adopting Jacobian transformations, we find that $d\alpha/d\tau = (da/dt)\times(t_0/a_0)$ and $de/d\tau = (de/dt)\times t_0$.
    For readability, we substitute the eccentricity factors with $F(e)=1+ (121/304)e^2$ and $G(e)=1+ (73/24)e^2 + (37/96)e^4$. 
    With these changes, we can now express Equations~\ref{eq:de_dt} and \ref{eq:da_dt} in non-dimensional form:
    \begin{eqnarray}
        \frac{de}{d\tau} &=& -\frac{19}{12}\frac{1}{\alpha^4}e\frac{F(e)}{(1-e^2)^{5/2}} \\
        \frac{d\alpha}{d\tau} &=& - \frac{1}{\alpha^3} \frac{G(e)}{(1-e^2)^{7/2}} 
    \end{eqnarray}

    These equations still exhibit the steep gradient at small $\alpha$ and the singularity at $\alpha=0$. To remove the singularity at $\alpha=0$, we make the substitution $s = -\ln(\alpha)$ (or $\alpha = \exp(-s)$). Since the separation monotonically decreases with time, we transform our parameterization so that $s$ is the independent variable instead of time, which allows for a finer resolution near the moment of merger. Specifically, we can define $a_\mathrm{contact}$ based on the Schwarzschild radius, or some other limit at arbitrarily small separations, and thus define $s_\mathrm{contact} = -\ln{(a_{\rm contact} / a_0)}$. Additionally, we apply a similar transformation of eccentricity into $\ln$-space, $e = \exp(l)$, to resolve small eccentricity with increased resolution and to avoid overstepping into small, non-physical negative eccentricities. After transforming, our coupled differential equations appear as:
    \begin{eqnarray}
        \frac{dl}{ds} &=& -\frac{19}{12} \left(1-e^2\right)\frac{F(e)}{G(e)} \label{eq:dl/ds}\\
        \frac{d\tau}{ds} &=& \exp({-4s}) \frac{(1-e^2)^{7/2}}{G(e)}. \label{eq:dtau/ds}
    \end{eqnarray}
    
    The price paid for this transformation is that $t_\mathrm{max}$ can no longer be the end condition for a solver, because $\tau$ has become a dependent variable. This issue is addressed with modern numerical integrators that use root-finding algorithms while solving to identify ``events''. In particular, we set up the integration, so the independent variable $s$ ranges from 0 (the value for the system's current separation) to the user defined $s_\mathrm{contact}$. If $\tau$ ever reaches $\tau_{\rm max} = t_\mathrm{max}/t_0$, the integration can be stopped at the requested time or continue till merger. 

    Unlike the original Equations \ref{eq:da_dt} and \ref{eq:de_dt}, the set of differential equations defined in Equations \ref{eq:dtau/ds} and \ref{eq:dl/ds} do not approach a singularity when approaching $a=0$ or $e=0$. We have implemented the updated form of these equations into a Python package that we make freely available for public use\footnote{\url{https://github.com/maxbriel/GW-integration}}. Our tests demonstrate that integration of the coupled set of Equations~\ref{eq:dl/ds} and \ref{eq:dtau/ds} is superior to the original equations from \citet{Peters+63} and \citet{Peters+64} for two reasons: First, when evolving a binary beyond its merger time, integrations using standard packages (e.g., the {\tt solve\_ivp} method within {\tt scipy.integrate}) are able to converge whereas integrating Equations~\ref{eq:da_dt} and \ref{eq:de_dt} do not (a warning is thrown and the integrator exhibits a failed flag). Convergence of the integrator means the merger time can be numerically identified within user-defined tolerances. Second, integration of the forms shown in Equations~\ref{eq:dl/ds} and \ref{eq:dtau/ds} requires significantly fewer function calls, speeding up the numerical integration itself; in our tests we find a 60\% to 70\% reduction.

    For many applications, integrating the standard equations from \citet{Peters+63} and \citet{Peters+64} until integration failure provides sufficient accuracy. However, there are cases where integration convergence is required. As such we have implemented this method into the binary population synthesis code {\tt POSYDON} \citep{Fragos+23, Andrews+25} for calculating the orbital evolution of compact object binaries. We caution that if a more precise merger time is necessary, higher-order approximations that extend the \citet{Peters+63} and \citet{Peters+64} equations, such as those found by \citet{Junker+92}, \citet{Gair+06}, or \citet{Zwick+20}, may be required. It may be possible to extend our approach to these results, but such an adaptation is beyond the scope of this work.
     
\begin{acknowledgments}
    This approach was motivated by a bug found by Bruno Vizzone in the binary population synthesis code {\tt POSYDON}. We thank him for taking the time to investigate and report this bug. We additionally thank Seth Gossage and Abhishek Chattaraj for their review of the code related to this bug. MMB is supported by the Swiss National Science Foundation (CRSII5\_213497). JJA acknowledges support for Program number (JWST-AR-04369.001-A) provided through a grant from the STScI under NASA contract NAS5-03127.

\end{acknowledgments}

\begin{contribution}
    MMB led the development of the method and numerical implementation. JJA helped with the method development. Both authors equally contributed to writing the manuscript.
\end{contribution}

\software{{\tt scipy} \citep{2020SciPy-NMeth}}

\bibliography{integrator}{}
\bibliographystyle{aasjournalv7}

\end{document}